\documentclass[12pt]{article}
\usepackage{amsmath,amssymb}
\usepackage{graphics}
\usepackage[dvipdfm]{graphicx}
\usepackage{color}
\usepackage{ulem}

\begin{document}
\baselineskip=16pt
\begin{titlepage}
\begin{flushright}
{\small OU-HET 716/2011}\\
\end{flushright}
\vspace*{1.2cm}

\begin{center}

{\Large\bf 
Vacuum stability in neutrinophilic Higgs doublet model
} 
\lineskip .75em
\vskip 1.5cm

\normalsize
{\large Naoyuki Haba} 
and
{\large Tomohiro Horita}

\vspace{1cm}

{\it Department of Physics, 
 Osaka University, Toyonaka, Osaka 560-0043, 
 Japan} \\

\vspace*{10mm}

{\bf Abstract}\\[5mm]
{\parbox{13cm}{\hspace{5mm}
%

A neutrinophilic Higgs model has tiny vacuum expectation value 
 (VEV), 
 which can naturally explain tiny masses of neutrinos. 
There is a large energy scale hierarchy 
 between a VEV of 
 the neutrinophilic Higgs doublet and that of usual 
 standard model-like Higgs doublet. 
In this paper 
 we at first analyze vacuum structures of 
 Higgs potential in
 both supersymmetry (SUSY) and non-SUSY
 neutrinophilic Higgs models,   
 and next 
 investigate 
 a stability of this VEV hierarchy 
 against radiative corrections.  
We will show that 
 the VEV hierarchy
 is stable against radiative corrections 
 in both Dirac neutrino and Majorana 
 neutrino scenarios in both SUSY and non-SUSY neutrinophilic 
 Higgs doublet models.

}}

\end{center}

\end{titlepage}

\section{Introduction}

The recent neutrino oscillation experiments
 gradually reveal a
 structure of 
 lepton sector\cite{Strumia:2006db, analyses}.
However, from the theoretical point of view, 
 smallness of neutrino mass is still a mystery and 
 it is one of the
 most important clues to find new physics beyond the
 standard model (SM). 
A lot of ideas have been suggested to explain 
 the smallness of neutrino masses comparing to those of quarks and
 charged leptons. 
How about considering a possibility that the 
 smallness of the neutrino masses 
 is originating from an extra Higgs
 doublet with a tiny vacuum expectation value (VEV). 
This idea is that 
 neutrino masses are much smaller than other fermions 
 because the origin of them comes from different VEV of different
 Higgs doublet, and then 
 we do not need  
 extremely tiny neutrino Yukawa coupling constants. 
This kind of model is so-called 
 neutrinophilic Higgs doublet
 model
\cite{Ma}-\cite{HS2}, 
 where a neutrinophilic Higgs take 
 a VEV of ${\mathcal O}(0.1)$ eV in
 Dirac neutrino scenario\cite{Nandi,WWY,Davidson:2009ha,Davidson:2010sf}, while 
 a VEV of ${\mathcal O}(1)$ MeV in Majorana neutrino scenario 
 with TeV-scale seesaw\cite{Ma,MaRa,Ma:2006km,HabaHirotsu,HabaTsumura,HS1,HS2}.  
The non-supersymmetric (non-SUSY) neutrinophilic 
 Higgs doublet model is sometimes called 
 $\nu$THDM. 
The (collider) phenomenology in $\nu$THDM 
 is interesting, 
 since a charged Higgs boson 
 is almost originated from the
 extra neutrinophilic Higgs doublet
 and its couplings to neutrinos
 are not small. 
The 
 characteristic signals 
 of the $\nu$THDM could be detected 
 at LHC and ILC experiments\cite{Davidson:2010sf,HabaTsumura}. 
Not small neutrino Yukawa couplings in the $\nu$THDM 
 can also make  
 low energy thermal 
 leptogenesis work\cite{HS1}. 
The SUSY version of neutrinophilic Higgs doublet model 
 have been suggested in Refs.\cite{HS1, HS2}, 
 where a thermal leptogenesis in a 
 low energy scale works without gravitino problem\cite{HS1, HS2}\footnote{Cosmological constraints were argued in Ref.\cite{Zhou:2011rc}, however,
a setup of them is different from usual neutrinophilic Higgs doublet models,
since it includes a light Higgs particle.}.

Anyhow, 
 a neutrinophilic Higgs model has tiny VEV,  
 and there is a large energy scale hierarchy 
 between a VEV of 
 the neutrinophilic Higgs doublet and that of usual 
 SM-like Higgs doublet. 
In this paper, 
 we at first analyze vacuum structures of 
 Higgs potential in
 both SUSY and non-SUSY
 neutrinophilic Higgs models,   
 and next 
 investigate 
 a stability of this VEV hierarchy 
 against radiative corrections.  
We will show that 
 the VEV hierarchy
 is stable against radiative corrections 
 in both Dirac neutrino and Majorana 
 neutrino scenarios in both SUSY and non-SUSY neutrinophilic 
 Higgs doublet models.


\section{$\nu$THDM}

Let us analyze vacuum structures of Higgs potential in 
 non-SUSY neutrinophilic Higgs model, i.e.,
 $\nu$THDM at first,
 and next investigate a stability of
 this VEV hierarchy against radiative corrections.

\subsection{Vacuum structure in tree-level potential}

We here overview the $\nu$THDM, 
 where we introduce
 a neutrinophilic Higgs doublet $\Phi_\nu$ and $Z_2$-parity 
 as follows. 
\begin{center}
\begin{tabular}{l||c|c} 
Fields & $Z_2$-parity & Lepton number \\ \hline
SM Higgs $\Phi$& + & 0 \\
neutrinophilic Higgs $\Phi_\nu $& $-$ & 0 \\
right-handed neutrino & $-$ & 1 \\
others & + & $\pm$ 1: leptons, 0: quarks \\
\end{tabular}
\end{center}
Yukawa interactions are given by 
\begin{equation}
\mathcal{L}_{\text{yukawa}} = y^u \bar{Q}_L \Phi U_R + y^d \bar{Q}_L
 \tilde{\Phi} D_R +y^l \bar{L} \Phi E_R + y^\nu \bar{L} \Phi_\nu N
+{\rm h.c.} ,
\end{equation}
where 
 $\tilde{\Phi} = i\sigma_2 \Phi$, and 
 generation indexes are omitted. 
Note that 
 the right-handed neutrino only couples with 
 $\Phi_\nu$ through the 
 Yukawa interaction, and this 
 is the origin of smallness of the neutrino masses.    
When we include Majorana mass of right-handed neutrinos 
 $\frac{1}{2} M\bar{N}^c N$,  
 this model becomes Majorana neutrino scenario 
 through the seesaw mechanism\cite{Type1seesaw}. 
A Higgs potential is given by 
\begin{multline}
V^{\nu\text{THDM}}= -m_1^2 \Phi^\dagger \Phi 
+ m_2^2 \Phi^\dagger_\nu \Phi_\nu 
- m_3^2( \Phi^\dagger \Phi_\nu+ \Phi_\nu^\dagger \Phi) 
+ \frac{\lambda_1}{2}( \Phi^\dagger \Phi)^2 
+ \frac{\lambda_2}{2}( \Phi_\nu^\dagger \Phi_\nu)^2  \\
+ \lambda_3 (\Phi^\dagger \Phi)(\Phi^\dagger_\nu \Phi_\nu)
+ \lambda_4 (\Phi^\dagger \Phi_\nu)(\Phi^\dagger_\nu \Phi)
+ \frac{\lambda_5}{2} [(\Phi^\dagger \Phi_\nu)^2 + (\Phi^\dagger_\nu \Phi)^2],
\end{multline}
where parameters are
 $m_1 \sim m_2 \sim {\mathcal O}(100)$ GeV 
 and 
 $\lambda_i \sim {\mathcal O}(1)$ $( i=1,...,5)$.
As for a magnitude of $|m_3^2|$, 
 we take (${\mathcal O}(10^{-0.5})$ GeV$)^2$
 for Majorana neutrino scenario, and 
 (${\mathcal O}(10^{-1})$ MeV$)^2$
 for Dirac neutrino 
 scenario. 
Notice that $\Phi$ has negative mass squared as $(-m_1^2)<0$.  
The Higgs doublets are assumed to be take real VEVs as 
 $\langle \Phi \rangle = (v_1, 0)^T$ and 
 $\langle \Phi_\nu \rangle = (v_2, 0)^T$,
 then, 
 stationary conditions are given by 
\begin{eqnarray}
0 = \frac{1}{2} \frac{\partial V^{\nu\text{THDM}}}
{\partial v_1} = -m_1^2 v_1 - m_3^2 v_2
 + \lambda_1 v_1^3 + \hat\lambda v_1 v_2^2, 
\label{sc1} \\
0 = \frac{1}{2} \frac{\partial V^{\nu\text{THDM}}}
{\partial v_2} = m_2^2 v_2 - m_3^2 v_1
 + \lambda_2 v_2^3 + \hat\lambda v_1^2 v_2 , 
\label{sc2}
\end{eqnarray}
where 
 $\hat\lambda \equiv \lambda_3 +\lambda_4 + \lambda_5$. 
We sort the following three cases by magnitude relations between 
 $|v_1|$ and $|v_2|$. 
\begin{enumerate}
\item {$|v_2| \ll  |v_1|$ case:} 
This vacuum is what the $\nu$THDM wants to realize. 
The magnitudes of VEVs are given by 
\begin{equation}
|v_1| \simeq \sqrt{\frac{m_1^2}{\lambda_1}},\hspace{1ex}
 v_2 \simeq
 \frac{m_3^2 v_1}{m_2^2 + \hat\lambda v_1^2} , 
\label{vev1}
\end{equation}
and a potential height at the vacuum is given by 
\begin{eqnarray}
V^{\text{THDM}}_{|v_2| \ll  |v_1|}  
\simeq - m_1^2 v_1^2 + \frac{\lambda_1}{2}v_1^4 \notag 
\simeq - \frac{m_1^4}{2 \lambda_1} . 
\label{potential1}
\end{eqnarray}
\item$|v_1| \ll |v_2|$ case: 
This vacuum suggests $v_2(m_2^2 + \lambda_2 v_2^2)=0$ from  
 Eq.(\ref{sc1}), and thus, 
\begin{eqnarray}
v_2^2 = \begin{cases}
                  0 & \text{($m_2^2  >0$)} ,\\
                  -\frac{m_2^2}{\lambda_2} & \text{($m_2^2 < 0$)}.
               \end{cases}
\label{26}
\end{eqnarray}
The case of 
 $v^2_2 = 0$ contradicts $|v_1| \ll |v_2|$. 
Another case of $v_2^2 =  -m_2^2/\lambda_2$ suggests 
 the value of $v_1$ as 
 $v_1^2 = m_3^2 v_2^2 / (-m_1^2 + \hat\lambda v_2^2)^2 > 0$, 
 where a 
 potential height is given by 
\begin{eqnarray}
V^{\nu\text{THDM}}_{|v_2| \gg  |v_1|}  
\simeq - \frac{m_2^4}{2 \lambda_2} . 
\label{potential1}
\end{eqnarray}
\item $|v_1| \sim |v_2|$ case: 
Neglecting tiny parameter $m_3^2$, 
 the stationary conditions Eqs.(\ref{sc1}) and (\ref{sc2}) become 
\begin{eqnarray}
-m_1^2 v_1 + \lambda_1 v_1^3 + \hat\lambda v_1 v_2 ^2 =0, \\
m_2^2 v_2 + \lambda _2 v_2^3 + \hat\lambda v_1^2 v_2 = 0. 
\end{eqnarray}
Then, VEVs 
 are given by 
\begin{eqnarray}
v_1^2 \simeq - \frac{\lambda_2 m_1^2 + \hat\lambda m_2^2}{\hat\lambda^2 - \lambda_1 \lambda_2},\hspace{1ex} v_2^2 &\simeq& \frac{\hat\lambda m_1^2 + \lambda_1 m_2^2}{\hat\lambda^2 - \lambda_1 \lambda_2} , \label{vev2}
\end{eqnarray}
and the potential height at the vacuum is estimated as 
\begin{eqnarray}
V^{\nu\text{THDM}}_{v_1 \sim v_2} 
&\simeq& \frac{\lambda_1 m_2^4 + \lambda_2 m_1^4 + 2\hat\lambda m_1^2
 m_2^2}{2(\hat\lambda^2 -\lambda_1 \lambda_2)}. 
 \label{potential2}
\end{eqnarray}
\end{enumerate}

Notice that the $\nu$THDM wants to realize 
 the vacuum in Eq.(\ref{vev1}) so that 
 this vacuum at
 $|v_2| \ll |v_1|$ should be a global minimum. 
Conditions of 
 $V^{\nu\text{THDM}}_{|v_2| \ll |v_1|} < V^{\nu\text{THDM}}_{|v_1| \sim |v_2|}$
 or 
 $V^{\nu\text{THDM}}_{|v_2| \gg |v_1|} < V^{\nu\text{THDM}}_{|v_1| \sim |v_2|}$
 suggest 
\begin{equation}
\hat\lambda^2 =  (\lambda_3+ \lambda_4 +\lambda_5)^2 > 
\lambda_1 \lambda_2. 
\label{stable}
\end{equation}
This is a necessary condition for $v_1 \gg v_2$ to be 
 the global minimum, and an additional condition 
 $-\frac{m_1^4}{2\lambda_1} < -\frac{m_2^4}{2\lambda_2}$
 makes the 
 vacuum  
 true global minimum. 
 For the potential to be bounded from below\cite{HabaHirotsu,Ginzburg:2007jn}, quartic terms must satisfy
\begin{equation}
 \sqrt{\lambda_1 \lambda_2} >  - ( \lambda_3 + \lambda_4 \pm \lambda_5),\;\; \sqrt{\lambda_1 \lambda_2} >  - \lambda_3, \;\;
  \lambda_1>0,\;\; \lambda_2 >0. 
 \label{bound}
 \end{equation}
These are the conditions of bounded below of the Higgs potential. 
We can show that 
 a case of 
 $\hat\lambda <0$ 
 cannot satisfy  the 
 global minimum condition. 
Therefore, 
 only a case of 
 $\hat\lambda >0$ can satisfy 
 the global minimum condition. 
Thus, in order for 
 the desirable vacuum $v_1 \gg v_2$ to be the global minimum, 
 a condition
 \begin{equation}
 0 < \sqrt{\lambda_1 \lambda_2} < \hat\lambda,\;\;\;  \sqrt{\lambda_1 \lambda_2} > -(\lambda_3 + \lambda_4 - \lambda_5),\;\;\; \lambda_1, \lambda_2 >0
 \end{equation}
is needed. 
 
Next, let us estimate a curvature (mass squared) at each vacuum,  
 which is given by 
\begin{eqnarray}
M_{ij}^2 =
 \frac{1}{2} \frac{\partial ^2 V^{\text{THDM}}}{\partial v_i \partial
 v_j}
 =
\begin{pmatrix}
-m_1^2 +3 \lambda_1 v_1^2 + \hat\lambda v_2^2 & -m_3^2 +2 \hat\lambda v_1 v_2 \\
-m_3^2 +2 \hat\lambda v_1 v_2 & m_2^2 +3 \lambda_2 v_2^2 + \hat\lambda v_1^2
\end{pmatrix}.
\end{eqnarray}
Then, the eigenvalue equation (eigenvalue: $x$) is given by  
\begin{eqnarray}
&& x^2 -(-m_1^2+m_2^2+(3\lambda_1 + \hat\lambda)v_1^2 +(3\lambda_2 + \hat\lambda)v_2^2 ) x -m_1^2 m_2^2 -m_3^4 + 3 \hat\lambda(\lambda_1 v_1^4 + \lambda_2 v_2^4) \notag \\
&& + (3\lambda_1 m_2^2 - \hat\lambda m_1^2)v_1^2  - (3\lambda_2 m_1^2 -
 \hat\lambda m_2^2)v_2^2 + 3(3 \lambda_1 \lambda_2 - \hat\lambda^2)v_1^2 v_2^2 +
 4m_3^2 \hat\lambda \lambda_1\lambda_2  =0,  
\label{ee}
\end{eqnarray}
and we can 
 estimate the curvature for above three cases. 
\begin{enumerate}
\item $|v_1| \gg |v_2|$ case: The eigenvalue equation in Eq.(\ref{ee})
 becomes 
\begin{eqnarray}
 x^2 -(-m_1^2+m_2^2+(3\lambda_1 + \hat\lambda)v_1^2  ) x + 3
  \hat\lambda\lambda_1 v_1^4 + (3\lambda_1 m_2^2 - \hat\lambda m_1^2)v_1^2
  -m_1^2 m_2^2 \simeq 0.  
\end{eqnarray}
By using Eq.(\ref{vev1}), 
 it becomes 
\begin{eqnarray}
(x - 2m_1^2)(x - \hat\lambda v_1^2 + m_2^2) = 0 ,
\end{eqnarray}
which means 
\begin{eqnarray}
x = 2m_1^2,\;\;\; \hat\lambda v_1^2 + m_2^2 .
\end{eqnarray}
Thus, 
 $m_1^2 >0$ and $\hat\lambda m_1^2 + \lambda_1 m_2^2$ must be needed for 
 $x > 0$. 
\item $|v_1| \ll |v_2|$ case: 
Using $v_2^2 = - \frac{m_2^2}{\lambda_2}$ in Eq.(\ref{26}), 
 the eigenvalue equation in Eq.(\ref{ee}) becomes 
\begin{eqnarray}
(x + 2m_2^2)(x - (\hat\lambda v_2^2 - m_1^2))=0. 
\end{eqnarray}
Thus, the solution is given by 
\begin{eqnarray}
x = -2m_2^2,\;\;\; \hat\lambda v_2^2 - m_1^2,
\end{eqnarray}
which means 
 $m_2^2 < 0, \hat\lambda m_2^2 + \lambda_2 m_1^2 < 0$ for 
 $x > 0$. 
\item $|v_1| \sim |v_2|$ case: 
By neglecting $m_3^2$ and using Eq.(\ref{vev2}), 
 the eigenvalue equation in Eq.(\ref{ee}) becomes 
\begin{eqnarray}
x^2 -2(\lambda_1 v_1^2 + \lambda_2 v_2^2)x -4(\hat\lambda^2 - \lambda_1
 \lambda_2) v_1^2 v_2^2  = 0, 
\end{eqnarray}
which means two eigenvalues $x_1,x_2$ should satisfy 
\begin{eqnarray}
&& x_1 + x_2 =  2(\lambda_1 v_1^2 + \lambda_2 v_2^2), 
\;\; x_1 x_2 = -4(\hat\lambda^2 - \lambda_1 \lambda_2) v_1^2 v_2^2.
\end{eqnarray}
Since positive $x_1,x_2$ give positive $x_1 + x_2,x_1 x_2$, 
 the positive curvature condition at this vacuum 
 is given by 
\begin{eqnarray}
\lambda_1 v_1^2 + \lambda_2 v_2^2 > 0 ,\label{eigen1} \\
-(\hat\lambda^2 - \lambda_1 \lambda_2) v_1^2 v_2^2 > 0 .\label{eigen2}
\end{eqnarray}
Thus, 
 $\hat\lambda^2 - \lambda_1 \lambda_2 < 0$, 
 is a positive curvature condition at the vacuum 
 of $|v_1| \sim |v_2|$. 
\end{enumerate}
The squared masses of the charged Higgs and of the pseudoscalar must be also positive. These conditions are equivalent to
\begin{eqnarray}
m_2^2 +\lambda_2 v_2^2 + \lambda_3 v_1^2 > 0 \label{ch} \\
m_2^2 + \lambda_2 v_2^2 + (\lambda_3 + \lambda_4 - \lambda_5) v_1^2 >0. \label{pseudo}
\end{eqnarray}

Summarizing conditions for the vacuum we want, 
at first, $\hat{\lambda}^2 - \lambda_2 \lambda_2$ must be positive for the vacua of $|v_1| \gg |v_2|$ and $|v_1| \ll |v_2|$ to be lower than that of $|v_1| \sim |v_2|$, and $-\frac{m_1^4}{2\lambda_1} < -\frac{m_2^4}{2\lambda_2}$ makes the vacuum of $|v_1| \gg |v_2|$  the global minimum.
Note that $\hat{\lambda}$ must be also positive to be consistent with the conditions of the potential bounded from below. 
Next, positive curvature conditions are $m_2^2> 0$ or $\hat\lambda m_1^2 + \lambda_1 m_2^2> 0$ with $m_2^2 < 0$.
Finally, positive curvature of the charged Higgs and the pseudoscalar components require $m_2^2  + \lambda_3 v_1^2 > 0$ and $m_2^2 + (\lambda_3 + \lambda_4 - \lambda_5) v_1^2 >0$ at $|v_1| \gg |v_2|$.
In Table 1, 
 we show which vacuum becomes the global minimum 
 depending on signs of $m_2^2$, 
 $\hat\lambda$, and $\hat\lambda^2 - \lambda_1 \lambda_2$. 
\begin{table}[!h]
\begin{center}
\newcommand{\bhline}[1]{\noalign{\hrule height #1}}
 \begin{tabular}{c|c|c||c|c||c|c||c|c}
 & ($m_2^2$, $\hat\lambda$) & $\hat\lambda^2 - \lambda_1 \lambda_2$
  &\multicolumn{2}{c||}{$|v_1| \gg |v_2|$}
 & \multicolumn{2}{c||}{$|v_1| \sim |v_2|$}
 & \multicolumn{2}{c}{$|v_1| \ll |v_2|$} \\ \hline
& & & GM  & PC & GM& PC & GM & PC \\ \hline
(1) & (+,+)& +  & \checkmark & \checkmark &  &  &  \checkmark &  \\ \hline
(2) & (+,+)& $-$  &  & \checkmark &  \checkmark & \checkmark &  \\ \hline
(3) & ($-$,+) & +  & \checkmark &  (a) & & & \checkmark & (b)\\ \hline
(4) & ($-$,+) & $-$  & & (a) & \checkmark & \checkmark & &  (b)    \\ \hline
(5) & (+,$-$) & $-$  &  & (a)  & \checkmark &\checkmark & \\ \hline
(6) & ($-$,$-$) & $-$   & & &\checkmark &  \checkmark & \\ 
  \end{tabular}
  \caption{
Six cases which satisfy conditions in Eq.(\ref{bound}), (\ref{ch}) and (\ref{pseudo}).
Here GM means 
 ``Global Minimum'' and PC means ``Positive Curvature'', and  
 $\checkmark$ in GM (PC) says each vacuum can be the 
 global minimum (has positive curvature).  
(a) and (b) mean that the positive curvature requires  
 conditions of   
 (a): $\hat\lambda m_1^2 + \lambda_1 m_2^2> 0$ and 
 (b): $-\hat\lambda m_2^2 - \lambda_2 m_1^2 > 0$, 
 respectively. 
Two simultaneous 
 $\checkmark$ in GM 
 means 
 $|v_1| \gg |v_2|$ ($|v_1| \ll |v_2|$) vacuum becomes the global minimum 
 when
  $-\frac{m_1^4}{2\lambda_1} < -\frac{m_2^4}{2\lambda_2}$ 
  ($-\frac{m_1^4}{2\lambda_1} > -\frac{m_2^4}{2\lambda_2}$). 
}
\end{center}
\end{table}

Can 
 a ``local minimum'' 
 at $|v_2| \ll |v_1|$ 
in (2), (4) and (5)
 be our vacuum? 
It might be possible
 if a life time of the local minimum 
 is long enough. 
There is a transition process from 
 the local minimum at $|v_2| \ll |v_1|$
 to the global minimum at $|v_1| \sim |v_2|$.
 Its transition probability of tunneling rate suggests the life time is
much shorter than an age of our universe, since
a ``distance" and a ``height" of wall between the local and global minimums
are both $\mathcal{O}$(100) GeV with $\mathcal{O}$(1) couplings of $\lambda_i$ in Higgs potential.
So, unfortunately, 
 the local minimum cannot 
 be our vacuum. 
Therefore, 
 in the $\nu$THDM,
 we must use the suitable parameter setup 
 as (1) or (3) with
 $-\frac{m_1^4}{2\lambda_1} < -\frac{m_2^4}{2\lambda_2}$.

Before closing this subsection, we comment on recent analyzes of vacuum structure in general THDM. For example, in Ref.\cite{Ginzburg:2007jn}, they investigated the vacuum instability of charge and/or CP breakings at tree level. 
As for so-called Inert Doublet Model (IDM) \cite{Gustafsson:2011wg}, it has exact $Z_2$-symmetry with $m_3^2 =0$.
This Inert Doublet does not couple with any matter fermions, which is crucial difference from our model.

\subsection{Stability against radiative corrections}

Now we are in a position to investigate 
 the stability of the VEV hierarchy $|v_2| \ll |v_1|$
 against radiative corrections. 
First of all, 
 we should remind that 
 the small magnitude of $|m_3^2|$ 
 plays a crucial role for generating the tiny 
 VEV of $|v_2| (\ll |v_1|)$. 
Its smallness is guaranteed against radiative corrections,
 since it is the ``soft'' breaking mass parameter of the 
 $Z_2$-symmetry. 
As noted in Ref.\cite{Davidson:2009ha}, the radiative correction to this parameter is expected to be logarithmic.
For analyses of the vacuum stability, 
 we should use Coleman-Weinberg 
 type 1-loop effective 
 potential\cite{Coleman:1973jx}, and 
 analyze 
 the stability of the VEV hierarchy. 
This 1-loop effective potential contains infinite number of irrelevant
 operators with zero-momentum Higgs fields in the external lines, and
 is calculated by a summation of them. 
However, for the investigation
 of stability of the VEV hierarchy, it is enough for us to pick up only
 diagrams which have external lines of mixture of 
 $\Phi$ and $\Phi_\nu$. 
Furthermore, 
 we should notice that, 
 when one $\Phi_\nu$ is added in the external 
 lines, a coefficient of the effective operator 
 should have  
 suppression factor, $|v_2/m_{1,2}|$. 
Thus, we investigate diagrams which have 
 only one $\Phi_\nu$ in the external lines. 

At first, 
 we focus on marginal 
 operators in the effective potential. 
The most dangerous marginal operator
 for the instability of the VEV hierarchy
 is $\lambda_6 |\Phi^2| (\Phi^{\dagger} \Phi_{\nu})$ (+h.c.), 
 which is induced from diagrams in Fig.1 (a) and (b). 
\begin{figure}[h]
\centering
\includegraphics[width=16cm,bb=0 0 1021 320, clip]{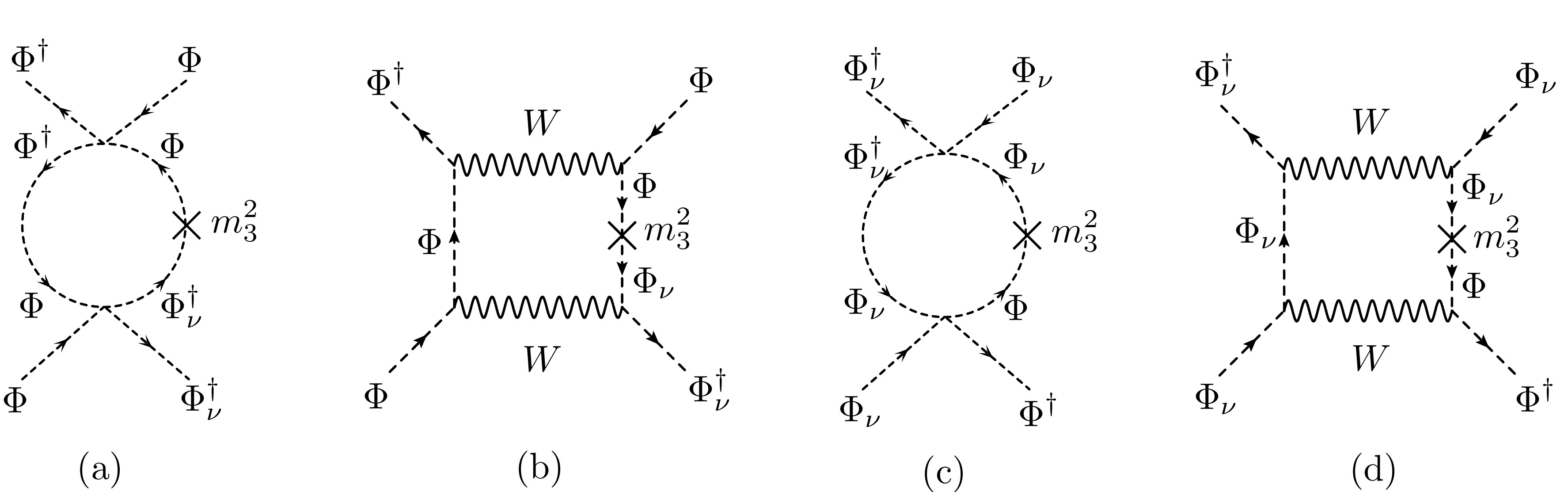} \\
Fig.1: $Z_2$-violating 1-loop diagrams.
\end{figure}
It is because this operator 
 breaks 
 $Z_2$-parity and
 induces 
 linear term of $v_2$, 
 which might possibly destroy  
 the VEV hierarchy.  
Here we note that Fig.1 (a) and (b) are 
 only 1-loop diagrams which induce   
 $\lambda_6 |\Phi^2| (\Phi^{\dagger} \Phi_{\nu})$ (+h.c.).  
Neither lepton nor quark 1-loop diagrams 
 contribute $\lambda_6$ due to the $Z_2$-parity,  
 since one additional external 
 $\Phi_{\nu}$ needs one additional 
 right-handed neutrino
 propagator inside a loop 
 which requires one more $\Phi_\nu$. 
Fig. 1 (c) and (d) induce  
 another $Z_2$-parity 
 violating operator, 
 $\lambda_7 |\Phi_{\nu}^2| (\Phi^{\dagger} \Phi_{\nu})$ (+h.c.).  
Figure 1 (a) ((c)) is expected to dominate (b) ((d)) because of $|\lambda_i|^2 \gg g_2^4$, so $\lambda_6$ and $\lambda_7$ are estimated as
\begin{eqnarray}
\lambda_6 &\sim& -\frac{3\lambda_1 \lambda_5}{4 \pi^2}
  \frac{m_3^2}{(m_2^2 - m_1^2)^2} 
 \left( m_2^2 - m_1^2 + m_2^2 \ln\frac{|m_1^2|}{|m_2^2|} \right) ,\\
\lambda_7 &\sim&  \frac{3\lambda_2 \lambda_5}{4
 \pi^2}  \frac{m_3^2}{(m_2^2 - m_1^2)^2} \left( m_2^2 - m_1^2 + m_1^2
					   \ln\frac{|m_1^2|}{|m_2^2|}
					  \right). 
\end{eqnarray}
Taking into account all irrelevant operators  
 which have 
 only one $\phi_\nu$ in the effective operator, 
 correction for $|\lambda_6|$ might be 
 of order
 $\frac{3}{2\pi^2}|\frac{m_3^2}{m_{1,2}^2}|\log|\frac{v_1}{v_2}|$ 
 at most. 
This correction
 contributes the stationary condition of 
 $v_2$ in Eq.(\ref{sc2}),
 and 
 modifies it as  
\begin{eqnarray}
0 = m_2^2 v_2 - m_3^2 v_1 + \lambda_2 v_2^3 +
 \hat\lambda v_1^2 v_2 +\frac{\lambda_6}{2} v_1^3 
+\frac{3\lambda_7}{2} v_1 v_2^2.
\end{eqnarray}
Remind again that 
 tiny VEV of $|v_2| (\ll |v_1|)$ is originated from 
 tiny term of 
 $m_3^2 v_1$. 
Thus, an induced term from the radiative correction of 
 $\frac{\lambda_6}{2} v_1^3$ must be smaller than 
 $m_3^2 v_1$ to preserve the VEV hierarchy.  
Actually, 
 the ratio of them is estimated as 
\begin{equation}
\left|\frac{\lambda_6 v_1^3}{2m_3^2 v_1}\right| \sim 
 \frac{3}{4\pi^2}\log\left|\frac{v_1}{v_2}\right|
\end{equation}
 at most.  
This means that the order of $|v_2|$ is not changed but its factor might
be modified about  0.8 (2)
by the radiative corrections in Majorana (Dirac) neutrino scenario.
This magnitude comes from a maximal (may be over-) estimation, and
anyhow, the orders of VEVs are not changed.
(Actually, this modification becomes much smaller about $\mathcal{O}$(1)\%,  if we
use Higgs self-couplings of $\mathcal{O}$(0.1).)
Thus, the VEV hierarchy itself is stable against radiative corrections.
As for higher-loop effects, they are at least suppressed by an additional
loop-factor $\frac{1}{16 \pi^2}$, and we cannot find any diagrams which have larger
contribution than above diagrams. 
Therefore, the VEV hierarchy itself is stable against
radiative corrections, and we can conclude radiative corrections do
not destroy the VEV hierarchy in both Dirac and Majorana neutrino
scenarios.

\section{SUSY neutrinophilic Higgs doublet model}

In this section, we analyze vacuum structures of Higgs potential
 in the SUSY neutrinophilic Higgs doublet model at first,
 and next investigate a stability of
 this VEV hierarchy against radiative corrections.

\subsection{Vacuum structure in tree-level potential}

The SUSY neutrinophilic Higgs doublet model has  
 four Higgs doublets\cite{HS1, HS2}, and   
 the superpotential is given by 
\begin{eqnarray}
\mathcal{W} = && y^u \bar{Q}^L H_u U_R + y^d \bar{Q}_L H_d D_R + \bar{L}
 H_d E_R + y^{\nu} \bar{L} H_{\nu} N \nonumber \\
&& + \mu H_u H_d  + \mu' H_{\nu} H_{\nu'} + \rho H_u H_{\nu'} + \rho' H_{\nu} H_d,
\label{WW}
\end{eqnarray}
where 
 $H_{\nu}$ gives Dirac neutrino masses and  
 $H_{\nu'}$ does not couple with any matters. 
Note that
 $H_u$ and $H_d$ are usual MSSM Higgs doublets. 
This superpotential is for Dirac neutrino scenario,
 and 
 Majorana neutrino scenario can be realized 
 when Majorana mass of right-handed neutrinos $M N^2$ 
 is included in Eq.(\ref{WW}). 
The $Z_2$-parity assignment of the fields
 is shown in the following table. 
\begin{center}
\begin{tabular}{l||c|c}
Fields & $Z_2$-parity & Lepton number \\ \hline
MSSM Higgs doublets $H_u, H_d$ & + & 0 \\ \hline
neutrinophilic Higgs doublets $H_{\nu}, H_{\nu'}$ & $-$ & 0 \\ \hline 
right-handed neutrino $N$ & $-$ & 1 \\ \hline
others & + & $\pm$ 1:leptons, 0:quarks
\end{tabular}
\end{center}
Note that $Z_2$-parity 
 is softly broken by $\rho, \rho'$, where 
 $|\rho|, |\rho'| \ll |\mu|, |\mu'|$. 
The Higgs potential is given by 
\begin{eqnarray}
 V &=& (|\mu|^2 +|\rho|^2) H_u^\dag H_u + (|\mu|^2+|\rho'|^2) H_d^\dag H_d 
      + (|\mu'|^2 +|\rho'|^2) H_{\nu}^\dag H_{\nu} + (|\mu'|^2+|\rho|^2) H_{\nu'}^\dag H_{\nu'}  \nonumber \\
  && + \frac{g_1^2}{2} \left( H_u^\dag \frac{1}{2} H_u - H_d^\dag\frac{1}{2} H_d 
     + H_{\nu}^\dag \frac{1}{2} H_{\nu} - H_{\nu'}^\dag \frac{1}{2}H_{\nu'} \right)^2  \nonumber \\
  && + \sum_a \frac{g_2^2}{2} \left( H_u^\dag \frac{\tau^a}{2} H_u + H_d^\dag\frac{\tau^a}{2} H_d 
     + H_{\nu}^\dag \frac{\tau^a}{2} H_{\nu} + H_{\nu'}^\dag \frac{\tau^a}{2}H_{\nu'} \right)^2  \nonumber \\
  && - m_{H_u}^2 H_u^\dag H_u  + m_{H_d}^2 H_d^\dag H_d 
     + m_{H_\nu}^2 H_{\nu}^\dag H_{\nu}+ m_{H_{\nu'}}^2 H_{\nu'}^\dag H_{\nu'} \nonumber \\
  && + B \mu H_u \cdot H_d + B' \mu' H_{\nu}\cdot H_{\nu'}
 + \hat{B} \rho H_u \cdot H_{\nu'} +
 \hat{B}' \rho' H_{\nu}\cdot H_{d}\nonumber\\
&& + \mu^* \rho H_d^\dag H_{\nu'}+\mu^* \rho' H_u^\dag H_{\nu}+
 \mu'^* \rho' H_{\nu'}^\dag H_{d}+\mu'^* \rho H_\nu^\dag H_{u}
 + \text{h.c.} ,
\end{eqnarray}
where $\tau^a$ and dot mean 
a generator and cross product of $SU(2)$, respectively. 
$\hat{B}'\rho'$ ($\hat{B}\rho$) 
 corresponds to $m_3^2$ in the non-SUSY 
 $\nu$THDM, and its smallness plays a 
 crucial role of generating tiny VEVs 
 of $H_{\nu, \nu'}$.  
The magnitude of $|\hat{B}'\rho'|$ ($|\hat{B}\rho|$) 
 is 
 (${\mathcal O}(10^{-0.5})$ GeV$)^2$
 for Majorana neutrino scenario, and 
 is 
  (${\mathcal O}(10^{-1})$ MeV$)^2$
 for Dirac neutrino 
 scenario. 
We assume $(-m^2_{H_u})<0$ for the suitable electroweak symmetry breaking 
 and real VEVs as    
\begin{equation}
\langle H_u \rangle = \begin{pmatrix} 0 \\ v_u  \end{pmatrix}, \langle
	      H_d \rangle = \begin{pmatrix} v_d \\ 0 \end{pmatrix},
					\langle H_{\nu} \rangle =
					 \begin{pmatrix} 0 \\ v_{\nu}
					 \end{pmatrix}, \langle H_{\nu'}
						   \rangle =
						   \begin{pmatrix}
						   v_{\nu'} \\ 0
						   \end{pmatrix}. 
\end{equation}
By taking $\mu,\rho, B$-parameters to be real  
 and denoting 
 $M_u^2 \equiv |\mu|^2 +|\rho|^2- m_{H_u}^2 (<0)$, 
 $M_d^2 \equiv |\mu|^2 +|\rho'|^2+ m_{H_d}^2 (>0)$, 
 $M_\nu^2 \equiv |\mu'|^2 +|\rho'|^2- m_{H_u}^2 (>0)$, and 
 $M_{\nu'}^2 \equiv |\mu'|^2 +|\rho|^2+ m_{H_d}^2 (>0)$, 
 the stationary conditions are given by 
\begin{eqnarray}
&&0 = \frac{1}{2} \frac{\partial V}{\partial v_u} 
   = M^2_u v_u + \frac{1}{4} (g_1^2 + g_2^2) v_u (v_u^2 - v_d^2 
     + v_{\nu}^2 - v_{\nu'}^2) - B \mu v_d -  \hat{B} \rho v_{\nu'} + (
     \mu \rho' + \mu' \rho ) v_{\nu}, \label{ssc1} \nonumber \\
&&0 = \frac{1}{2} \frac{\partial V}{\partial v_d} 
   = M^2_d v_d - \frac{1}{4}(g_1^2 + g_2^2) v_d (v_u^2 - v_d^2 
     + v_{\nu}^2 - v_{\nu'}^2) -  B \mu v_u -  \hat{B}' \rho' v_{\nu} + ( \mu \rho +  \mu' \rho' ) v_{\nu'} ,\label{ssc2} \nonumber \\
&&0 = \frac{1}{2} \frac{\partial V}{\partial v_{\nu}} 
   = M^2_{\nu} v_{\nu} + \frac{1}{4}(g_1^2 + g_2^2) v_{\nu} (v_u^2 - v_d^2 
     + v_{\nu}^2 - v_{\nu'}^2) -  B' \mu' v_{\nu'} -  \hat{B}' \rho' v_d + (\mu \rho' +  \mu' \rho ) v_u ,\label{ssc3} \nonumber \\
&&0 = \frac{1}{2} \frac{\partial V}{\partial v_{\nu'}} 
   = M^2_{\nu'} v_{\nu'} - \frac{1}{4}(g_1^2 + g_2^2) v_{\nu'} (v_u^2 - v_d^2 
     + v_{\nu}^2 - v_{\nu'}^2) -  B' \mu' v_{\nu} - \hat{B} \rho v_u + (
     \mu \rho +  \mu' \rho' ) v_d . \nonumber \label{ssc4} 
\end{eqnarray}
Let us investigate the vacuum structure 
 with a 
 parametrization of 
 $v_u = v \sin{\beta} \cos{\gamma}, 
 v_d = v \cos{\beta} \cos{\gamma}, 
 v_\nu = v \sin{\beta_\nu} \sin{\gamma}, 
 v_{\nu'} = v \cos{\beta_{\nu}} \sin{\gamma}$.   
At first, we focus on the vacuum which neutrinophilic 
 Higgs doublet model requires, i.e.,  
 $|v_u|, |v_d| \gg |v_{\nu}|, |v_{\nu'}|$. 
This condition  
 induces the usual MSSM relations for $v_u, v_d$ as 
\begin{eqnarray}
M^2_u - \frac{1}{4} (g_1^2 + g_2^2)v^2 \cos{2\beta} - B \mu \cot{\beta}
 \simeq 0 , \;\;
M^2_d + \frac{1}{4} (g_1^2 + g_2^2) v^2 \cos{2\beta} - B \mu \tan{\beta}
 \simeq 0, \notag
\end{eqnarray}
which means 
\begin{eqnarray}
&& v^2 \simeq  \frac{2}{g_1^2 + g_2^2} \left( \frac{M_u^2 -
					M_d^2}{\cos{2\beta}} -(  M_u^2 +
					M_d^2) \right) , \;\;\;\;\;
 \sin{2\beta} \simeq  \frac{2B \mu}{M_u^2 + M_d^2 }. 
\end{eqnarray}
They induce tiny $v_{\nu}, v_{\nu'}$ through tiny 
 $\rho,\rho'$ as 
\begin{align}
v_{\nu} = \frac{ \left[ M_{\nu'}^2 - \frac{1}{4} (g_1^2 + g_2^2)(v_u^2 - v_d^2 ) \right] \left[ \hat{B}' \rho' v_d - (\mu \rho' +  \mu' \rho ) v_u \right] + B' \mu' \left[ \hat{B} \rho v_u - ( \mu \rho +  \mu' \rho' ) v_d \right] }
{\left[ M_{\nu}^2 + \frac{1}{4} (g_1^2 + g_2^2)(v_u^2 - v_d^2 ) \right]
\left[ M_{\nu'}^2 - \frac{1}{4} (g_1^2 + g_2^2)(v_u^2 - v_d^2 ) \right] 
- B'^2 \mu'^2}  \label{334}, \\
v_{\nu'} = \frac{ \left[ M_{\nu}^2 + \frac{1}{4} (g_1^2 + g_2^2)(v_u^2 - v_d^2)  \right] \left[ \hat{B} \rho v_u - (\mu \rho +  \mu' \rho' ) v_d \right]  + B' \mu' \left[  \hat{B}' \rho' v_d - ( \mu \rho' + \mu' \rho ) v_u \right] }
{ \left[ M_{\nu}^2 + \frac{1}{4} (g_1^2 + g_2^2)(v_u^2 - v_d^2 )\right]
\left[ M_{\nu'}^2 - \frac{1}{4} (g_1^2 + g_2^2)(v_u^2 - v_d^2 ) \right] 
- B'^2 \mu'^2} .
\end{align}
At this vacuum, the potential height is estimated as 
\begin{eqnarray}
V \simeq v^2 (M_u^2 \sin^2{\beta} + M_d^2 \cos^2{\beta} - 2B \mu \cos{\beta} \sin{\beta}) + \frac{1}{8}(g_1^2 + g_2^2)v^4 \cos^2{2 \beta}.
\end{eqnarray}

Next, we study the vacuum at 
 $|v_u|, |v_d| \sim |v_\nu|, |v_{\nu'}|$. 
Where, by neglecting both $\rho$ and $\rho'$, 
 the stationary conditions
 become 
\begin{eqnarray}
M^2_u v_u + \frac{1}{4} (g_1^2 + g_2^2) v_u (v_u^2 - v_d^2 + v_{\nu}^2 - v_{\nu'}^2) - B \mu v_d =0   ,\\
M^2_d v_d - \frac{1}{4}(g_1^2 + g_2^2) v_d (v_u^2 - v_d^2  + v_{\nu}^2 - v_{\nu'}^2) -  B \mu v_u  =0 ,\\
M^2_{\nu} v_{\nu} + \frac{1}{4}(g_1^2 + g_2^2) v_{\nu} (v_u^2 - v_d^2 + v_{\nu}^2 - v_{\nu'}^2) -  B' \mu' v_{\nu'} =0  ,\\
M^2_{\nu'} v_{\nu'} - \frac{1}{4}(g_1^2 + g_2^2) v_{\nu'} (v_u^2 - v_d^2
 + v_{\nu}^2 - v_{\nu'}^2) -  B' \mu' v_{\nu}  =0 .
\end{eqnarray}
It is easy to show that 
 only $v_\nu=v_\nu'=0$ can satisfy 
 the stationary conditions 
 in $D$-flat direction of $v_\nu=v_\nu'$.

Numerical analyzes show that the vacuum at $v_\nu=v_\nu'=0$ is the
global minimum in suitable parameter regions of $|B'|, |\mu'|= {\mathcal O}(10^2)$ GeV and positive $M_\nu, M_{\nu'}= {\mathcal O}(10^2)$ GeV.
This result is originated from 
 an initial setup that 
 only soft mass squared of $H_u$ 
 is negative.  
(See, case (1) of Table 1 in $\nu$THDM.)
Similarly, 
 we can show that there is no vacuum at 
 $|v_u|, |v_d| \ll |v_\nu|, |v_{\nu'}|$. 
Anyhow, the vacuum exits only at 
  $|v_u|, |v_d| \gg |v_\nu|, |v_{\nu'}|$, 
 which is the desirable vacuum in 
 the neutrinophilic Higgs doublet model.

\subsection{Stability against radiative corrections}

Let us investigate the stability 
 of the VEV hierarchy against radiative 
 corrections in the SUSY neutrinophilic 
 Higgs doublet model. 
As in non-SUSY case, we can estimate 1-loop radiative corrections in a
SUSY effective potential.
\begin{center}
\includegraphics[width=13cm,bb=0 0 1021 393,clip]{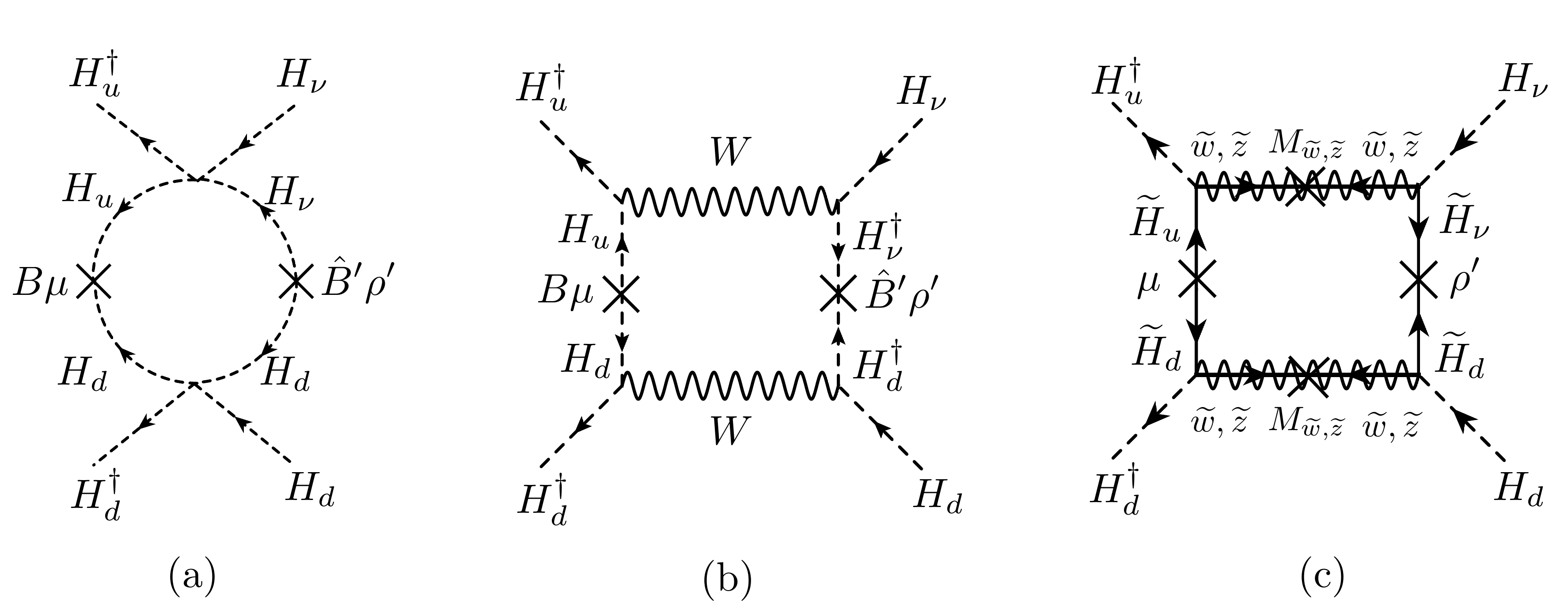}\\
Fig.2: $Z_2$-violating 1-loop diagrams in SUSY. 
\end{center}
The most dangerous marginal 
 operator in the effective potential is
 $\lambda' (H_u^{\dagger} H_{\nu})(H_d^{\dagger} H_d)$ (+h.c.),
 which is induced from 
 $Z_2$-violating diagrams in Figs.2 (a)$\sim$(c). 
The absolute value of $\lambda'$ is roughly estimated 
 as 
 $\frac{g_2^4}{8 \pi^2}|\frac{\hat{B}'\rho'}{m^2}|$
 at most,  
 where $m$ is a Higgs mass in a loop. 
Notice that neither (s)lepton nor (s)quark contribute 
 $\lambda'$ at 
 1-loop level 
 due to the $Z_2$-parity
 similarly in non-SUSY $\nu$THDM. 
It is because one additional external 
 $H_{\nu}$ needs one additional 
 right-handed neutrino propagator inside a loop, 
 which requires one more external $H_\nu$.  
Anyhow, this term modifies the stationary condition 
 of 
 $v_{\nu}$ in Eq.(\ref{ssc3}) as 
\begin{eqnarray}
0 = M^2_{\nu} v_{\nu} - \frac{1}{4}(g_1^2 + g_2^2) v_{\nu} \left[ v_u^2 - v_d^2 + v_{\nu}^2 - v_{\nu'}^2  +\frac{ 2\lambda'}{(g_1^2 + g_2^2)} \frac{v_u v_d^2}{v_{\nu}} \right] - \hat{B}' \rho' v_d + (\mu \rho' +  \mu' \rho ) v_u.
\label{34}
\end{eqnarray}
Taking into account all irrelevant operators  
 which have 
 only one $H_\nu$ in the effective operator, 
 correction for $|\lambda'|$ might be 
 of order
 $\frac{g_2^4}{4 \pi^2}|\frac{\hat{B}'\rho'}{m^2}
 |\log|\frac{v_{u,d}}{v_\nu}|$ at most. 
Remind that 
 tiny VEV of $v_\nu$ is originated from the small mass parameters of 
 $\hat{B}' \rho'$ as in Eq.(\ref{334}).   
Thus, in order to preserve the VEV hierarchy, 
 $|\frac{\lambda'}{2} v_u v_d^2|$ must be smaller than 
 $|\hat{B}' \rho' v_d|$ in Eq.(\ref{34}).   
And,
 this ratio is estimated as  
\begin{eqnarray}
\left|\frac{\lambda' v_u v_d^2}{2\hat{B}' \rho' v_d}\right| \sim
 \frac{g_2^4}{8 \pi^2}\left|\frac{v_uv_d}{m^2}\right|
 \log\left|\frac{v_{u,d}}{v_\nu}\right|. 
\end{eqnarray}
This value is too small to influence 
 the stationary 
 conditions in both 
 Dirac and Majorana neutrino scenarios.  
We can also show that higher-loop diagrams induce 
 smaller corrections due to 
 the loop suppression factors. 
Therefore, we can conclude 
 that the potential is stable against radiative corrections 
 in SUSY neutrinophilic Higgs doublet model.

\section{Summary}

A neutrinophilic Higgs model has tiny VEV,  
 which can naturally explain tiny masses of neutrinos. 
There is a large energy scale hierarchy 
 between a VEV of 
 the neutrinophilic Higgs doublet and that of usual 
 SM-like Higgs doublet. 
In this paper, 
 we have analyzed vacuum structures of 
 Higgs potential in
 both SUSY and non-SUSY
 neutrinophilic Higgs models,   
 and next 
 investigated 
 a stability of this VEV hierarchy 
 against radiative corrections.   
We have shown that 
 the VEV hierarchy 
  is stable against radiative corrections 
 in both Dirac neutrino and Majorana 
 neutrino scenarios in both SUSY and non-SUSY neutrinophilic 
 Higgs doublet models.

\vspace{1cm}

{\large \bf Note added}\\

\vspace{-.5cm}
\noindent
After preparing our submission of this paper, we notice a paper  
 \cite{Morozumi}, where  
 authors also analyzed the vacuum stability against 
 radiative corrections in the 
 non-SUSY $\nu$THDM with Dirac neutrino scenario. 
Their results are consistent with ours.
 They calculated the 1-loop effective potential and the quantum corrections to VEV hierarchy. 
 On the other hand, we estimated the most dangerous contributions to the VEV hierarchy and confirmed the stability also in SUSY and Majorana cases.


\vspace{1cm}

{\large \bf Acknowledgments}\\

\vspace{-.2cm}
\noindent
We thank K. Tsumura, O. Seto, and N. Maru  
 for useful and helpful discussions. 
This work is partially supported by Scientific Grant by Ministry of 
 Education and Science, Nos. 20540272, 20039006, and 20025004.


\end{document}